\documentclass[12pt]{iopart}
\usepackage{graphicx}
\begin{document}

\title[]{High pressure study of the normal and superconducting states of the layered pnictide oxide Ba$_{1-x}$Na$_x$Ti$_2$Sb$_2$O with x = 0, 0.10, and 0.15}
\author{M. Gooch$^1$, P. Doan$^2$, B. Lorenz$^1$,  Z. J. Tang$^2$, A. M. Guloy$^{2}$, and C. W. Chu$^{1,3}$}

\address{$^1$ TCSUH and Department of Physics, University of Houston, Houston, TX 77204, USA}
\address{$^2$ TCSUH and Department of Chemistry, University of Houston, Houston, TX 77204, USA}
\address{$^3$ Lawrence Berkeley National Laboratory, 1 Cyclotron Road, Berkeley, CA 94720, USA}

\begin{abstract}
Here we present a systematic study of the effects of pressure on the superconducting and spin/charge density wave (SDW/CDW) transitions of Ba$_{1-x}$Na$_x$Ti$_2$Sb$_2$O (x = 0, 0.10, and 0.15) by means of resistivity measurements. For x = 0 and 0.10, external pressure results in a decease of the SDW/CDW transition temperature T$_c$; however, no measurable change is observed for the x = 0.15. The pressure effect on the superconducting transition temperature is different for all three samples. For BaTi$_2$Sb$_2$O (x=0), T$_c$ increases significantly from 1.2 K at zero pressure to $\sim$ 2.9 K at 16.1 kbars. The 10 \% Na-doped sample shows an initial T$_c$ increase up to 4.2 K with pressure which saturates at higher pressure values. For higher Na concentrations (x=0.15), T$_c$ continuously decreases with increasing pressure.
\end{abstract}


\section{Introduction}
The initial discovery of superconductivity in iron pnictides by Kamihara \textit{et. al.} \cite{Kamihara08} continues to fuel research efforts, which has resulted in the discovery of superconductivity in many new layered transition metal-based pnictides and chalcogenides \cite{Greene10}. Both the iron pnictides and chalcogenides have a layered structure similar to that seen in the cuprates. More interestingly, superconductivity in many iron pnictides/chalcogenides seems to emerge through chemical doping in close proximity to a spin or charge density wave. This ordering is well understood for low dimensional metallic systems and emerges when there is nesting of the Fermi surface \cite{Lomer62,Wilson75,Whangbo91}. The ordered state of the parent compounds is antiferromagnetism (AF) in the copper oxide superconductors, a spin density wave (SDW) state in the iron pnictides, and a charge density wave (CDW) state in layered chalcogenides \cite{Chu10,Greene10,Norman11,Sipos08,Morosan06}. In the continuing search for new superconducting materials researchers are interested in layered compounds with different transition metals which also have electronic instabilities in their undoped parent compounds.

One family of compounds suggested for a new class of superconductors is Na$_2$Ti$_2$\textit{Pn}$_2$O which has been studied recently \cite{Adam90,Ozawa08,Johrendt11}.  Na$_2$Ti$_2$\textit{Pn}$_2$O has a layered structure, as well as, an electronic instability whose ordering temperature is T$_S$ = 330 K and T$_S$ = 120 K respectively for \textit{Pn} = As \cite{Brock95} and \textit{Pn} = Sb \cite{Adam90}. Previously performed band structure calculation by Pickett revealed a square fermi surface with a nesting feature \cite{Pickett98,deBiani98} which is also associated with a periodic lattice distortion \cite{Ozawa00}.  Much like the pnictides these planar titanium pnictide oxides are a member of a larger materials family including Na$_2$Ti$_2$\textit{Pn}$_2$O (\textit{Pn} = Sb, As), (SrF)$_2$Ti$_2$\textit{Pn}$_2$O (\textit{Pn} = Sb,As), and (SmO)$_2$Ti$_2$Sb$_2$O \cite{Adam90,Axtell97,Ozawa00,Ozawa01,Ozawa04,Liu09,Liu10,Brock95,Ozawa08}; however, unlike the iron pnictides where superconductivity was found for many cations and inter planar distances, none of the above materials were superconducting.

Wang \textit{et. al.} found that Na$_2$ in Na$_2$Ti$_2$As$_2$O could be completely replaced by Ba resulting in a new compound with similar Ti$_2$O layers and a SDW/CDW instability near T$_S$=220 K. Attempts to induce superconductivity in BaTi$_2$As$_2$O through doping or intercalation of different ions have not been successful, although a systematic lowering of the T$_S$ through the intercalation of Li$^+$ had been observed \cite{Wang10}. Superconductivity was finally found below 1.2 K in BaTi$_2$Sb$_2$O \cite{Yajima12}, and in the Na-doped Ba$_{1-x}$Na$_x$Ti$_2$Sb$_2$O \cite{Doan12}, with T$_c$'s up to 5.5 K. More recently, the mixed compounds BaTi$_2$(Sb$_{1-x}$Bi$_x$)$_2$O and BaTi$_2$(Sb$_{1-x}$Sn$_x$)$_2$O have also been found superconducting with relatively low values of T$_c$ \cite{Yajima13,Zhai13,Nakano13}.

BaTi$_2$Sb$_2$O has the lowest T$_S$=54 K and a T$_c$=1.2 K. Through Na$^+$ doping, T$_S$ continuously decreases while T$_c$ increases to a maximum of 5.5 K \cite{Doan12}.  BaTi$_2$Sb$_2$O shares the layered structure of other high T$_c$ superconductors with Ti$_2$O forming a square lattices, similar yet inverse to CuO$_2$ planes of the cuprates, with the positions of the transition metal and oxygen ions interchanged. First principle calculations have proposed different possible scenarios to explain the occurrence of superconductivity in BaTi$_2$Sb$_2$O and the Na-doped analogue. Subeki suggested that the high-temperature anomaly is due to the formation of a CDW, with conventional superconductivity arising from a strong electron-phonon coupling \cite{Alaska13}.  This is in contrast to Singh's proposal, derived form the assumption that the nesting property of the Fermi surface results in the formation of a SDW. In the later case, unconventional superconductivity is driven by magnetic fluctuations, with a pairing state of sign-changing s-wave symmetry \cite{Singh12}. Recent NMR and muon spin measurements are in favor of a commensurate CDW order at T$_S$ and an ordinary s-wave superconducting state below T$_c$ \cite{Kitagawa13,Rohr13}; however, further work is still warranted to elucidate the complex interplay between the SDW/CDW order and the superconductivity in Ba$_{1-x}$Na$_x$Ti$_2$Sb$_2$O and related compounds.

Therefore, to further investigate this system and to reveal the possible changes of the SDW/CDW and superconducting transition temperatures upon application of hydrostatic pressure we have conducted a systematic study of the resistivity of Ba$_{1-x}$Na$_x$Ti$_2$Sb$_2$O (x=0, 0.1, 0.15) as a function of pressure. The results show a correlation of the critical temperatures, T$_S$ and T$_c$, indicating a possible competition of the SDW/CDW and the superconducting phases.

\section{Experimental}
Ba$_{1-x}$Na$_x$Ti$_2$Sb$_2$O polycrystalline samples were synthesized as described previously \cite{Doan12}. All experimental handling of these air and moisture sensitive samples was performed within a purified Ar-atmosphere glovebox, with total H$_2$O and O$_2$ levels of $<$ 0.1 ppm.  To obtain dense samples for the pressure measurements, all powder samples were pressed into pellets and subsequently annealed. Prior to physical characterization, chemical and structural analyses were conducted to confirm single phase purity on all samples used for the high-pressure measurements. At ambient pressures the superconducting transitions, defined as the midpoint of the resistivity drop, for Ba$_{1-x}$Na$_x$Ti$_2$O is 2.1 K, 3.8 K, and 5.2 K for x = 0, 0.10, and 0.15 respectively. To gain insight into the effect of pressure on the normal and superconducting states, a four probe resistivity measurement was conducted with low-frequency (19 Hz) ac bridge LR 700 (Linear Research). Hydrostatic pressure up to 18 kbars were generated with a beryllium copper clamp cell, with Fluorinert FC70 as the pressure medium. A lead manometer was used to measure the pressure \textit{in situ} with the LR700 Inductance Bridge \cite{Chu74}.

\begin{figure}
\begin{center}
	\includegraphics[angle=0,width=6in]{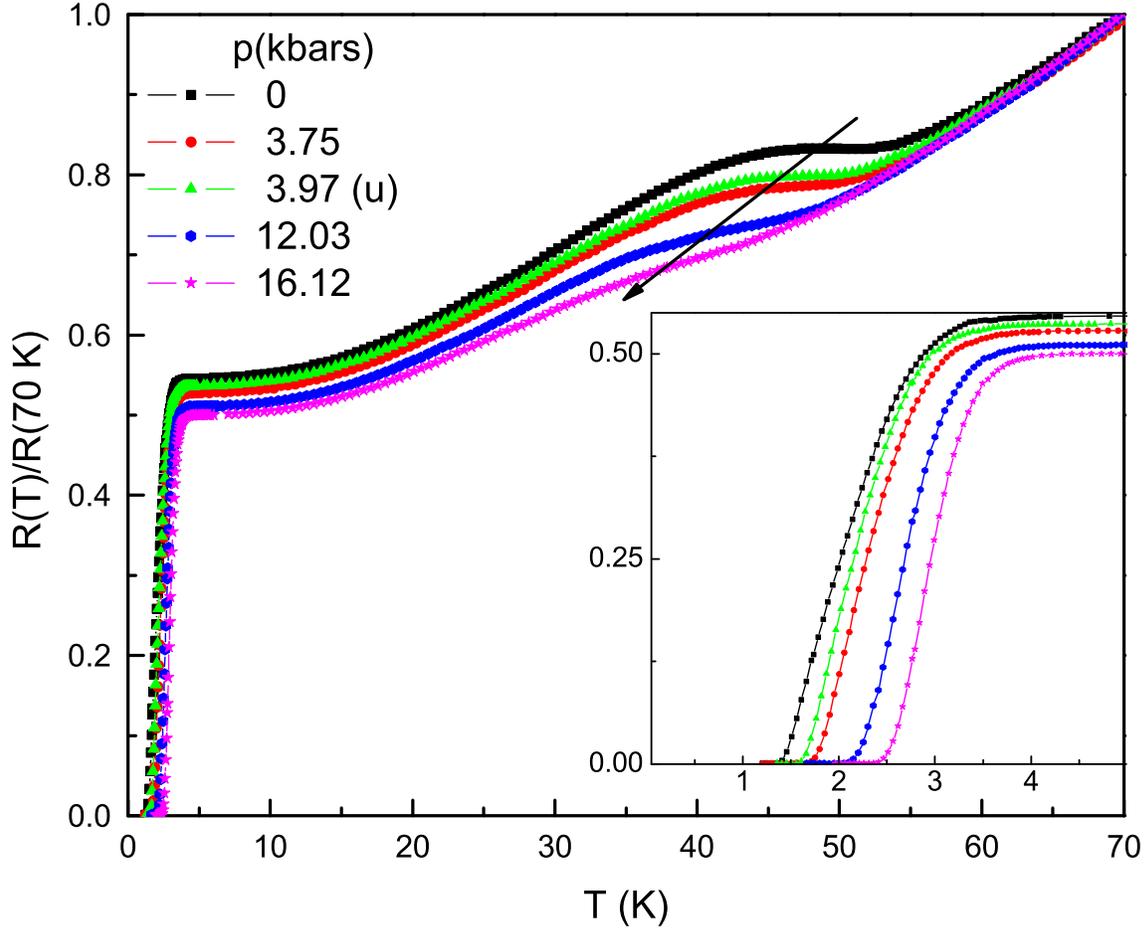}
	\caption{(Color online)  Normalized resistivity for BaTi$_2$Sb$_2$O under pressure, where u denotes unload pressure run and the inset shows the low temperature normalized resistivity.}
	\label{fig1}
\end{center}
\end{figure}

\section{Results and Discussions}
The undoped BaTi$_2$Sb$_2$O has the highest SDW/CDW transition temperature, as well as the sharpest resistivity anomaly at T$_S$ = 54 K, coupled with a low superconducting T$_c$ = 2.1 K as shown in Fig. ~\ref{fig1}. The superconducting transition is relatively broad with the onset of the resistivity drop at 3.5 K and zero resistance at $\sim$ 1.2 K, consistent with earlier reports \cite{Yajima12}. The broad superconducting transition can be related to the nature of polycrystalline samples and grain-grain coherence within the sample. With increasing pressure T$_S$ continually decreases linearly, Fig 4a, whereas the superconducting transition temperature increases and becomes sharper, as seen in the insert of Fig. ~\ref{fig1}. At the maximum pressure of 16.12 kbars, T$_c$ has increased from 2.1 K (p=0) to 2.9 K. To verify the stability of the sample the pressure cell was unloaded to lower pressure, the corresponding data are denoted in Fig. ~\ref{fig1} as (u).  T$_c$ shifts with a positive and linear pressure coefficient, (see Fig 4b), with a small pressure coefficient dlnT$_c$/dp $\sim$ 0.02 kbars$^{-1}$.  In contrast, the SDW/CDW phase is suppressed by pressure and T$_S$ decreases to $\sim$40 K. The decrease of T$_S$ is similar to the effect of Na$^+$ doping in Ba$_{1-x}$Na$_x$Ti$_2$Sb$_2$O, where with increased Na doping a broadening and decrease of the T$_S$ is observed. It should be noted that a similar suppression of T$_S$ in iron pnictides for R(O/F)FeAs and (Ba,Sr)$_{1-x}$K$_x$FeAs was attributed to the pressure induced increase in the hole density of the Fe$_2$As$_2$ layers \cite{Chu09}.

\begin{figure}
\begin{center}
	\includegraphics[angle=0,width=6in]{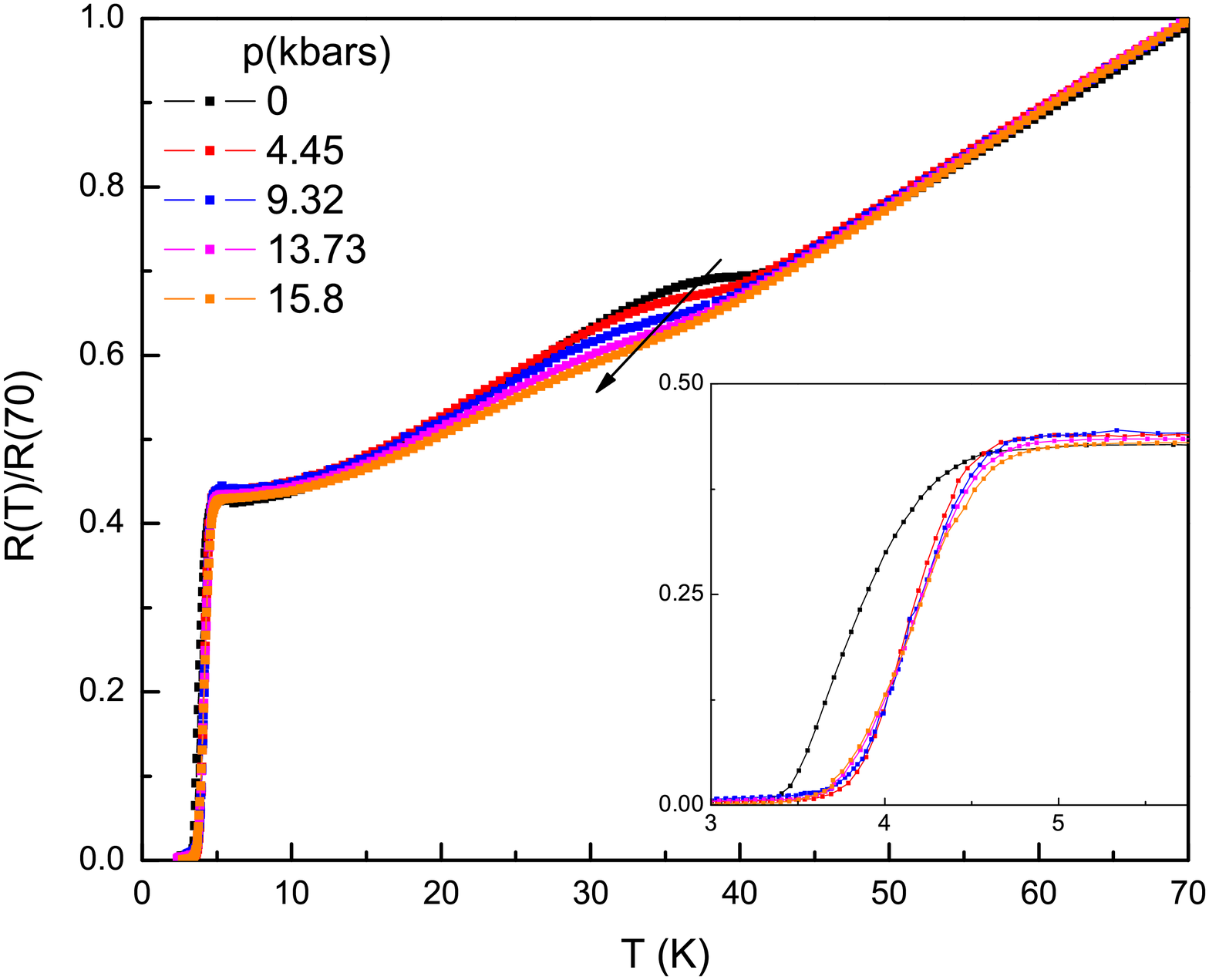}
	\caption{(Color online)  Normalized resistivity of Ba$_{0.90}$Na$_{0.10}$Ti$_2$Sb$_2$O under pressure with the inset showing the low temperature normalized resistivity..}
	\label{fig2}
\end{center}
\end{figure}

Ba$_{0.90}$Na$_{0.10}$Ti$_2$Sb$_2$O, much like the parent compound shows the coexistence of the SDW/CDW and superconductivity. At ambient pressure, Ba$_{0.90}$Na$_{0.10}$Ti$_2$Sb$_2$O exhibits considerably lower T$_S$ of 39 K and a higher T$_c$=3.8 K as shown in Fig.~\ref{fig2}. The T$_S$ decreases further and broadens with increasing pressure, similar to the parent compound. The superconducting transition sharpens, and initially increases up to 4.17 K in the low pressure range; however, with further increase of pressure up to 15.8 kbars, the T$_c$ remains almost constant within the measurement resolution (Fig. 4b). While both the parent compound and 10 $\%$ Na both share a decrease of the SDW/CDW anomaly with increasing pressure, the pressure shift of T$_c$ apparently depends greatly on the doping state.  As seen in Fig 4b, the relative pressure shift initially is comparable to the parent compound with dlnT$_c$/dp $\sim$ 0.02 kbars$^{-1}$.  However, with further increasing pressure T$_c$ remains constant within the measurement resolution. This is similar to what had been observed in iron pnictide superconductors \cite{Chu09}.

\begin{figure}
\begin{center}
	\includegraphics[angle=0,width=6in]{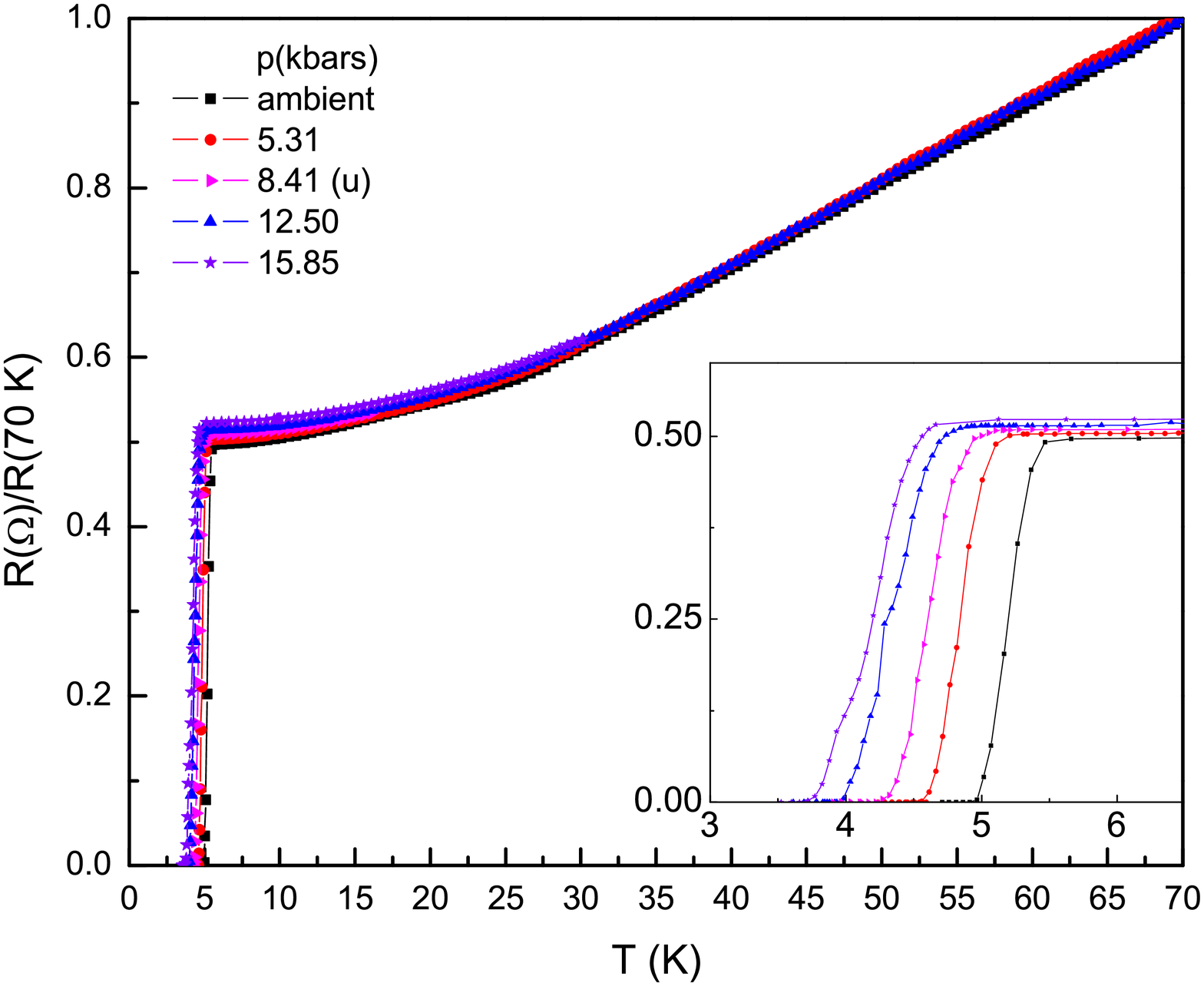}
	\caption{(Color online)  Normalized resistivity for Ba$_{0.85}$Na$_{0.15}$Ti$_2$Sb$_2$O under pressure, u denotes unload pressure run and the inset shows the low temperature normalized resistivity..}
	\label{fig3}
\end{center}
\end{figure}

\begin{figure}
\begin{center}
	\includegraphics[angle=0,width=6in]{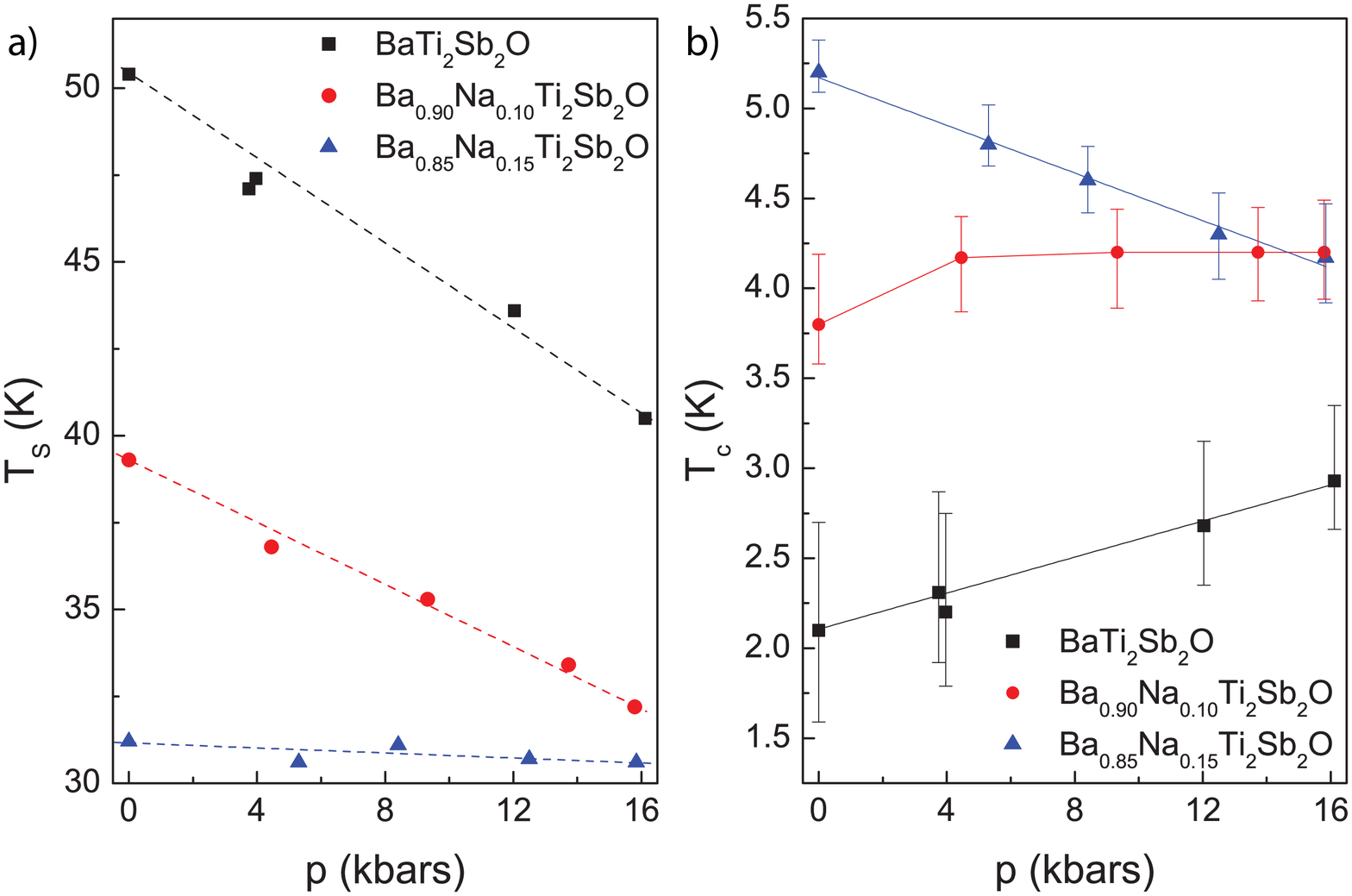}
	\caption{(Color online) (a) The shift of T$_S$ with pressure for BaTi$_2$Sb$_2$O, Ba$_{0.9}$Na$_{0.1}$Ti$_2$Sb$_2$O, and Ba$_{0.85}$Na$_{0.15}$Ti$_2$Sb$_2$O.  (b) The shift of T$_c$ with pressure for BaTi$_2$Sb$_2$O, Ba$_{0.9}$Na$_{0.1}$Ti$_2$Sb$_2$O, and Ba$_{0.85}$Na$_{0.15}$Ti$_2$Sb$_2$O.  The error bars denote the 90, 50, and 10 percent drop of the superconducting transition.}
	\label{fig4}
\end{center}
\end{figure}

Ba$_{0.85}$Na$_{0.15}$Ti$_2$Sb$_2$O is near the optimal doping state with the highest ambient pressure T$_c$ = 5.2 K (Fig.~\ref{fig3}). With increasing pressure, T$_c$ decreases linearly up to the maximum pressure of 15.85 kbars (see Fig. 4b). The superconducting transition broadens slightly with increasing pressure, possibly due to small stress induced by the solidification of the pressure medium, and T$_c$ dropped to 4.2 K at the highest pressure.  We also observe that the T$_c$ values measured upon loading and unloading the pressure cell fall on the same line (Figs.~\ref{fig3},\ref{fig4}) proving the stability of the sample in the pressure cycle. The decrease of T$_c$ with pressure coefficient of dlnT$_c$/dp = -0.01 kbars$^{-1}$, is similar to the pressure effect reported for the iron pnictides and cuprate superconductors in the optimally and overdoped regime. The pressure shifts for all three samples are shown in Fig. 4b. It is obvious that the pressure shift of T$_c$ in the Ba$_{1-x}$Na$_{x}$Ti$_2$Sb$_2$O system strongly depends on the doping state. However, the absolute value of T$_c$ in samples with different Na doping could not be increased by pressure above the maximum T$_c$ of 5.5 K observed at ambient pressure for the optimal doping \cite{Doan12}.

While the layered structure type of BaTi$_2$Sb$_2$O may be analogous to the high-T$_c$ cuprate superconductors, they differ by the Ti and O atoms that form the OTi$_{4/2}$ antiperovskite layers which are capped by Sb atoms in the Ti$_2$Sb$_2$O slabs.  Furthermore, the stacking of the Ti$_2$Sb$_2$O slabs in BaTi$_2$Sb$_2$O differ from those in Na$_2$Ti$_2$Pn$_2$O in that the nearest neighboring Ti$_2$Sb$_2$O slabs which are alternately stacked with Ba/Na atoms along the c axis, are oriented in an \textit{A-A} manner (\textit{A-B} in Na$_2$Ti$_2$Pn$_2$O).  The \textit{A-A} slab orientation results in Sb-Sb interlayer distances that are essentially nonbonding.  This is unlike the isotypical CeCr$_2$Si$_2$C type, where the Si$_2$ have short Si interlayer distances. From previous work it is known that as Na$^+$ is doped into the structure, it contracts the Ti-Ti distances within the layer, and results in a volume contraction \cite{Doan12}. However, the Na doping also affects the charge carrier concentration and the density of states at the Fermi energy, N(E$_F$). Within a rigid band model, Na$^+$ replacing Ba$^{2+}$ (hole doping) reduces the number of electrons in the Ti$_2$Sb$_2$O slab resulting in a decrease of E$_F$ and an increase of N(E$_F$), according to recent band structure calculations \cite{Singh12}. This can easily explain the increase of T$_c$ with doping if we assume a BCS-like mechanism for superconductivity where N(E$_F$) is a relevant parameter. It should be noted that there is evidence for a weak coupling BCS-like superconductivity in Ba$_{1-x}$Na$_{x}$Ti$_2$Sb$_2$O from recent NMR \cite{Kitagawa13}, heat capacity \cite{Gooch13}, and muon spin rotation measurements \cite{Rohr13}.

The effect of physical pressure on the superconducting state is more complex \cite{Lorenz05} and it cannot simply be explained by chemical pressure induced by Na doping. The strong doping dependence of the pressure coefficients observed in Ba$_{1-x}$Na$_{x}$Ti$_2$Sb$_2$O appears to be similar to the copper oxide and some iron pnictide superconductors \cite{Gooch08}. In those systems, dT$_c$/dp is positive in the underdoped case but negative above optimal doping. The sign change of dT$_c$/dp with doping was explained in cuprates by a pressure-induced charge transfer from the charge reservoir block to the active superconducting CuO$_2$ plane, thus following the generic dome shaped phase boundary of high-temperature copper oxide superconductors. In iron pnictide superconductors, the proximity of the superconducting state to a magnetic quantum critical point could be essential and the suppression of the spin density wave phase by pressure has to be considered to understand the pressure effect on T$_c$.

In Ba$_{1-x}$Na$_{x}$Ti$_2$Sb$_2$O the Ti$_2$Sb$_2$O layer is sandwiched between planes of Ba$^{2+}$/Na$^+$ ions. A pressure-induced charge transfer from Ba$^{2+}$ or Na$^+$ to the superconducting Ti$_2$O plane is very unlikely, however, a charge transfer between the Sb 5p bands and the Ti 3d bands may be considered. Although most of the Sb 5p bands lie well below E$_F$, a small fraction extends to the Fermi energy and hybridizes with the Ti 3d orbitals, forming the covalent Sb-Ti bond \cite{Singh12}. A pressure-induced transfer of electrons from Ti to Sb which is equivalent to hole doping into the Ti$_2$O plane cannot be ruled out. It would explain the increase of T$_c$ for the undoped and underdoped compounds as well as the decrease at higher doping. In addition, the suppression of the SDW/CDW state, which is apparently competing with the superconducting state, is expected to play an important role and to contribute to the stabilization of the superconducting state for low doping. At higher doping, the possible reduction of N(E$_F$), induced by a band broadening under pressure, can contribute to the T$_c$ decrease. There are certainly several competing mechanisms involved in the complex response of the superconducting state and its critical temperature to external pressure. 

More experimental and theoretical work is needed to completely understand the observed pressure effects. For example, calculations of the band structure and the electron-phonon interactions, similar to those conducted recently \cite{Singh12,Alaska13} can simulate the pressure effects by reducing the volume and relaxing the lattice coordinates with the goal to show the influence of pressure on the density of states as well as the electron-phonon coupling constants. Heat capacity experiments under pressure may also provide useful information about pressure-induced changes of the electronic state (N($E_F$)) and the lattice. The role of the density wave phase and the coexistence with superconductivity is another interesting topic that deserves further studies. Up to this point, there is no experimental evidence for a magnetically ordered state below T$_S$. Therefore, recent experiments have been in favor of a charge density wave order coexisting with superconductivity \cite{Kitagawa13,Rohr13}. This coexistence is unusual and calls for further investigations.

\section{Summary}
The pressure effects on the SDW/CDW and superconducting transitions of the system Ba$_{1-x}$Na$_x$Ti$_2$Sb$_2$O was investigated systematically for different doping levels, x=0, 0.1, 0.15. The SDW/CDW transition temperature decreases with applied pressure; however, the superconducting T$_c$ shows a more complex dependence on applied pressure and doping. An increase of T$_c$ with pressure is observed for x=0 and x=0.1, but T$_c$ decreases with pressure for the optimally doped case x=0.15. These results show an intricate relation between the superconducting and the SDW/CDW states. While the nature of the ordered SDW/CDW state has yet to be explored, it may have an important influence on the superconducting state, the pairing symmetry, and the gap function. The high-pressure data indicate that the superconductivity in Ba$_{1-x}$Na$_x$Ti$_2$Sb$_2$O possibly competes with the SDW/CDW order. The understanding of this order is crucial and more work is warranted to understand its nature and its relation to the superconducting state.

\ack
This work is supported in part by the T.L.L. Temple Foundation, the J. J. and R. Moores Endowment, the State of Texas through Texas Center for Superconductivity, the U.S. Air Force office of Scientific Research, and at LBNL through USDOE.  Support from the NSF (CHE-0616805) and the R. A. Welch Foundation (E-1297) is also gratefully acknowledged.
\section*{References}
\bibliographystyle{unsrt}

\end{document}